\documentclass[12pt]{article}
\catcode`\@=11
\@addtoreset{equation}{section}

\global\arraycolsep=2pt
\usepackage{mathrsfs,amsbsy,amssymb,latexsym,amsfonts,amsmath,cite}
\usepackage{graphicx,color}
\usepackage{physics}
\usepackage{mathtools}
\usepackage[normalem]{ulem}
\usepackage{caption}
\usepackage{here}

\usepackage{mathrsfs,amsbsy,amssymb,latexsym,amsfonts,amsmath,cite,bm}
\usepackage{graphicx,color}
\usepackage{mathtools}
\usepackage{enumerate}

\DeclareFontFamily{U}{BOONDOX-calo}{\skewchar\font=45 }
\DeclareFontShape{U}{BOONDOX-calo}{m}{n}{
  <-> s*[1.05] BOONDOX-r-calo}{}
\DeclareFontShape{U}{BOONDOX-calo}{b}{n}{
  <-> s*[1.05] BOONDOX-b-calo}{}
\DeclareMathAlphabet{\mathcalboondox}{U}{BOONDOX-calo}{m}{n}
\SetMathAlphabet{\mathcalboondox}{bold}{U}{BOONDOX-calo}{b}{n}
\DeclareMathAlphabet{\mathbcalboondox}{U}{BOONDOX-calo}{b}{n}

\allowdisplaybreaks

\begin{document}

\renewcommand{\thefootnote}{\fnsymbol{footnote}}

\begin{flushright}
RUP-24-22\\
STUPP-24-275
\end{flushright}
\vspace*{0.5cm}

\begin{center}
{\Large \bf  Oscillons in AdS space
}
\vspace*{1cm} \\
{\large
Takaaki Ishii$^{1}$\footnote{E-mail:~ishiitk$\_$at$\_$rikkyo.ac.jp}, 
Takaki Matsumoto$^{2,3}$\footnote{E-mail:~takaki-matsumoto$\_$at$\_$ejs.seikei.ac.jp},
Kanta Nakano$^{3}$\footnote{E-mail:~k.nakano.233$\_$at$\_$ms.saitama-u.ac.jp}, \\
Ryosuke Suda$^{3}$\footnote{E-mail:~r.suda.813$\_$at$\_$ms.saitama-u.ac.jp}
and Kentaroh Yoshida$^{3}$\footnote{E-mail:~kenyoshida$\_$at$\_$mail.saitama-u.ac.jp}} 
\end{center}

\vspace*{0.4cm}

\begin{center}
$^{1}${\it Department of Physics, Rikkyo University, \\
3-34-1 Nishi-Ikebukuro, Toshima-ku, Tokyo 171-8501, Japan}
\end{center}
\begin{center}
$^{2}${\it Seikei University, \\
3-3-1 Kichijoji-Kitamachi, Musashino-shi, Tokyo 180-8633, Japan}
\end{center}
\begin{center}
$^{3}${\it Graduate School of Science and Engineering, Saitama University, \\
255 Shimo-Okubo, Sakura-ku, Saitama 338-8570, Japan}
\end{center}

\vspace{1cm}

\begin{abstract}
We study oscillons in a real scalar field theory in a (3+1)-dimensional AdS space with global coordinates. The initial configuration is given by a Gaussian shape with an appropriate core size as in Minkowski spacetime. The solution exhibits a long lifetime. In particular, since the AdS space can be seen as a box, the recurrence phenomenon can be observed under suitable conditions. {In particular, as the AdS radius decreases, one can see a transition from a metastable oscillon to a stable oscillatory solution.} Finally, we discuss some potential applications of the oscillon in the context of AdS/CFT duality.
\end{abstract}

\setcounter{footnote}{0}
\setcounter{page}{0}
\thispagestyle{empty}

\newpage

\tableofcontents

\renewcommand\thefootnote{\arabic{footnote}}

\section{Introduction}

Oscillons are localized and long-lived excitations in a real scalar field theory with a self-interacting potential in $d+1$ dimensions (where $d$ is the number of spatial directions). It is well known that excitations in (1+1)-dimensional Minkowski spacetime may be stable due to Derrick's theorem \cite{Derrick}. In higher dimensions ($d>1$), there is no such theorem and the mechanism to stabilize excitations has been unknown. However, there may be localized and long-lived excitations, whose lifetime is extremely long but finite. They are oscillons.

\medskip 

The oscillons have some basic characteristics, which are summarized as follows. The oscillons are 
\begin{enumerate}
\item \quad stabilized by non-linear effect from non-trivial potential,  \vspace{-0.2cm}
\item \quad time-dependent, long-lived excitations with no topological charge,  \vspace{-0.2cm}
\item \quad spherically symmetric,  \vspace{-0.2cm}
\item \quad strongly sensitive to the initial configuration e.g., Gaussian profile.  \vspace{-0.2cm}
\end{enumerate}
The oscillons were originally recognized by Gleiser 
\cite{Gleiser:1993pt} and were theoretically grounded in a series of papers \cite{Copeland:1995fq,Gleiser:2008ty,Gleiser:2009ys}. However, it is fair to say that the fundamental mechanism to ensure the longevity of oscillons has been quite unclear, although we know that some specific initial configurations lead to the long-lived excitations phenomenologically. 

\medskip 

In this letter, we investigate oscillons in a real scalar field theory in a (3+1)-dimensional anti-de Sitter (AdS) space with global coordinates. If we take the Gaussian profile as the initial configuration, oscillons can be realized, just as in Minkowski spacetime. The oscillon behavior depends on the curvature radius $\ell$\,. In particular, since the AdS space can be seen as a box, a phenomenon of recurrence can be observed for suitable values of $\ell$\,. This may be worth noting because, although the system is non-integrable, the oscillons exhibit some integrable-ish properties.  {Furthermore, as the AdS radius decreases, one can see a transition from a metastable oscillon to a stable oscillatory solution.} 

\medskip 

This letter is organized as follows. In Section 2, we will introduce the setup to study oscillons in AdS. 
In Section 3, we perform numerical computations and show the existence of oscillons. In particular, the oscillons exhibit the recurrence phenomenon. {In addition, we find a transition from a metastable oscillon to a stable oscillatory solution.} Section 4 is devoted to the conclusion and discussion. {In Appendix A, we explain how to rescale for numerical computations. In Appendix B, towards including quantum corrections, we present an oscillon with parameters for which perturbative computations are valid. In Appendix C, we describe how to derive an inequality for the oscillon core size in the AdS case. Intriguingly, this can also be seen as a bound for possible values of the AdS radius.}

\section{Setup}

In this section, let us prepare the system we will analyze. We first introduce the classical action of a real scalar field theory with a self-interacting potential in a ($d$+1)-dimensional AdS space. Then the initial and boundary conditions to realize oscillons will be introduced by following the Minkowski case \cite{Copeland:1995fq}. Finally, the shell energy is defined so as to figure out the behavior of the oscillons and to track the radiations emitted from the shell.

\subsection{A real scalar field theory in AdS}

We consider a real scalar field theory in a $d+1$ dimensional global AdS space with the metric
\begin{align}
    ds^2 = g_{\mu\nu} dx^{\mu} dx^{\nu}
    =
    -\left(1+\frac{r^2}{\ell^2}\right)dt^2
    +\left(1+\dfrac{r^2}{\ell^2}\right)^{-1}dr^2
    +r^2d\Omega_{d-1}^2\,,
    \label{adsds2}
\end{align}
where $\ell$ is the curvature radius. The velocity of light $c$ is taken to be 1.

\medskip 

Suppose here that the scalar field $\phi$ is spherically symmetric. Then it depends only on $t$ and $r$ as $\phi=\phi(t,r)$\,. Then the classical action 
\begin{eqnarray}
 S\left[\phi\right]
    = 
    \int\!d^{d+1}x\, \sqrt{-g}\,\left[
    -\frac{1}{2}\,g^{\mu\nu}\,\partial_{\mu}\phi\,\partial_{\nu}\phi 
    - V(\phi) \right]\,
\end{eqnarray}
is reduced to  
\begin{align}
    S\left[\phi\right]
    = 
    V_{d-1}
    \int\!dt\!\int\! dr\, r^{d-1}\left[
    -\frac{1}{2}\,g^{tt}\,\partial_{t}\phi\,\partial_{t}\phi 
    -\frac{1}{2}\,g^{rr}\,\partial_{r}\phi\,\partial_{r}\phi 
    - V(\phi) \right]\,,
    \label{action}
\end{align}
where $V_{d-1}$ is the $(d-1)$-dimensional unit sphere volume and is given by 
\[
 V_{d-1} = \frac{2\pi^{d/2}}{\Gamma\left(d/2\right)}\,.  
\]
In the following, we will consider an asymmetric double well potential, 
\begin{align}
    V(\phi) = \frac{m^2}{2}\phi^2 - \frac{\alpha_0}{3}\phi^3 + \frac{\lambda}{4}\phi^4\,, 
    \label{potential}
\end{align}
where $m$ is mass, and $\lambda$ and $\alpha_0$ are real positive parameters.

\medskip

To carry out numerical calculations, we need to make the parameters dimensionless. Although the curvature radius $\ell$ is commonly utilized to make the coordinates dimensionless, we will use the mass $m$ here\footnote{Conventionally, the mass term seems to play an important role. However, even in the massless case, an oscillon solution was constructed in \cite{Dorey} although the scalar potential is represented by a geometric series and essentially has an infinite number of terms. In this letter, we will not consider the massless case.}. For details of the rescaling, see Appendix A. As a result, the dimensionless Lagrangian is given by
\begin{align}
    \mathcal{L}_\mathrm{rescaled}\left[\phi\right]
    =
    \frac{1}{\lambda} V_{d-1} \,r^{d-1}
    \left[
    -\frac{1}{2}\,g^{tt}\,
    \partial_t\phi\,\partial_t\phi
    -\frac{1}{2}\,g^{rr}\,
    \partial_r\phi\,\partial_r\phi
    -\frac{1}{2}\phi^2
    +\frac{\alpha}{3}\phi^3
    -\frac{1}{4}\phi^4
    \right]\,,
    \label{Lagrngian_rescaled}
\end{align}
Note here that all the quantities are now dimensionless and different from the original ones in (\ref{action})\,. We have abbreviated the form (\ref{dimless}) by omitting the hats. 

\medskip 

The equation of motion is given by 
\begin{align}
    -f^{-1}\,\partial_{t}^{2}\phi + f\,\partial_{r}^{2}\phi + \frac{d-1}{r}f\,\partial_{r}\phi +  \frac{2r}{L^2}\partial_{r}\phi
    - \phi + \alpha\phi^2 - \phi^3 = 0\,,
\end{align}
where the function $f(r)$ is defined as 
\begin{align}
    f(r) \equiv 1 + \frac{r^2}{L^2}\,.
\end{align}
Taking $L\,\rightarrow\,\infty$\,, we obtain the equation of motion in Minkowski spacetime.

\subsection{Oscillon ansatz}

In order to realize oscillons in our setup, let us suppose the same initial and boundary conditions as in the case of Minkowski spacetime \cite{Copeland:1995fq}, which are the following four conditions:
\begin{align}
    &\partial_t\phi(t=0,\,r) = 0\,,\label{bc1}\\
    &\partial_r\phi(t,\,r=0) = 0\,,\label{bc2}\\
    &\phi(t,\,r=\infty) = 0\label{bc4}\,, \\ 
    &\phi(t=0,\,r) = {\rm e}^{-r^2/R_0^2}\,.\label{bc3}
\end{align}
The first condition \eqref{bc1} requires that the initial velocity is zero for all values of $r$\,. The second and third ones \eqref{bc2} and \eqref{bc4} impose the regularity at $r=0$ and $r=\infty$\,, respectively. The final \eqref{bc3} supposes that the initial shape is Gaussian, where $R_0$ is the oscillon core size at $t=0$\,. It is known phenomenologically that this Gaussian shape gives rise to longevity in the case of Minkowski spacetime \cite{Copeland:1995fq}\footnote{In \cite{Copeland:1995fq}, initial configurations with other shapes like a tanh function are also discussed. However, it seems likely that the Gaussian is the best for longevity}.

\subsection{Shell energy}

It is convenient to introduce the notion of the shell energy in order to capture the oscillon time evolution and to track the radiations emitted from the oscillon. 

\medskip 

The shell energy is defined as 
\begin{align}
    E_{\rm s}(t) \equiv \int_0^{R_\mathrm{s}}\!\!dr\,\mathcal{E}\left[\phi\right]\,, 
    \label{Es_def}
\end{align}
where the integrand is the energy density given by 
\begin{align}
    \mathcal{E}\left[\phi\right]
    =
    \frac{1}{\lambda}\,V_{d-1}\,r^{d-1}\left[\frac{1}{2}\,f^{-1}\,\partial_{t}\phi\,\partial_{t}\phi 
    + \frac{1}{2}\,f\,\partial_r\phi\,\partial_r\phi
    + V(\phi)
    \right]\,,
\end{align}
and $R_\mathrm{s}$, measuring the shell size, is taken to be sufficiently large in comparison to the oscillon core size $R_0$\,. By definition, the shell energy contains all the energy of the oscillon solution at $t=0$\,. As time goes on, radiation goes out of the shell and the shell energy decreases (but it is useful to see a  behavior of $E_{\rm s}$ intrinsic to the AdS space). Hence $E_{\rm s}$ is basically time dependent.

\medskip

As a matter of course, the total energy 
\begin{align}
    E_{\rm tot} \equiv \int_0^{\infty}\!\!dr\,\mathcal{E}\left[\phi\right] 
    \label{E}
\end{align}
is conserved due to the time translation symmetry of the classical action.

\section{Oscillons in AdS}

In this section we shall perform numerical calculations and find oscillon solutions. Some characteristic properties of the oscillons as well as some peculiar behaviors in the AdS space are presented. 

\subsection{Conformal map}

Before going into the details, it is convenient to use a conformal map for our numerical computation to compactify the radial direction to a finite interval. The conformal transformation for the radial direction is given by 
\begin{equation}
    r = L\,\tan{\theta}\,. 
\end{equation}
Then the radial direction $0\leq r <\infty$ is compactified to $0\leq \theta <\pi/2$\,.

\medskip

After performing the transformation, the differential equation to be studied is given by 
\begin{align}
    -\partial^2_\tau\phi
    +\partial^2_\theta\phi
    +\frac{d-1}{\sin{\theta}\cos{\theta}}\partial_\theta\phi
    -\frac{L^2}{\cos^2{\theta}}\left(\phi-\alpha\phi^2+\phi^3\right) = 0\,.
\label{diff-dimless}
\end{align}
Here, we have rescaled the original time $t$ as $t\equiv L\,\tau$\,. The Gaussian profile \eqref{bc4} in the initial conditions is similarly rewritten as 
\begin{align}
    \phi(t=0,\,r) = \exp\left(-\frac{L^2\tan^2{\theta}}{R_0^2}\right)\,.
\end{align}
The shell energy $E_{\rm s}$ is also given by 
\begin{align}
    E_{\rm s}(t) \equiv \int_0^{\theta_\mathrm{s}}\!\!d\theta\,\frac{L}{\cos^2{\theta}}\,\mathcal{E}\left[\phi\right]\,, 
    \label{Es}
\end{align}
where $\theta_{\rm s}$ is defined by $R_{\rm s} = L \tan\theta_{\rm s}$ and the integrand is 
\begin{align}
    \mathcal{E}\left[\phi\right]
    =
    \frac{1}{\lambda}\,V_{d-1}\,L^{d-3}\,\cos^2{\theta}\,\tan^{d-1}\!{\theta}\left[\frac{1}{2}
    \partial_\tau\phi\partial_\tau\phi
    + \frac{1}{2}
    \partial_\theta\phi\partial_\theta\phi
    + \frac{L^2}{\cos^2{\theta}}\,V(\phi)
    \right]\,.
\end{align}
The total energy can be obtained by replacing $\theta_{\rm s}$ with $\pi/2$\,.

\medskip 

In the following, we will consider the case with $d=3$ and $\lambda=1$\,. 

\subsection{Oscillon longevity} 

We solve the differential equation (\ref{diff-dimless}) using a central finite difference scheme with second-order accuracy in both space and time. In the conformal coordinates, the spatial direction is discretized as uniform grids consisting of $15707$ (i.e.~$d\theta=1.0\times10^{-4}$) spatial grid points and a time step of $dt=3.0\times10^{-7}$\,. We then observe a long-lived localized excitation.

\medskip 

\begin{figure}[htbp]
    \centering
    \includegraphics[width=12 cm]{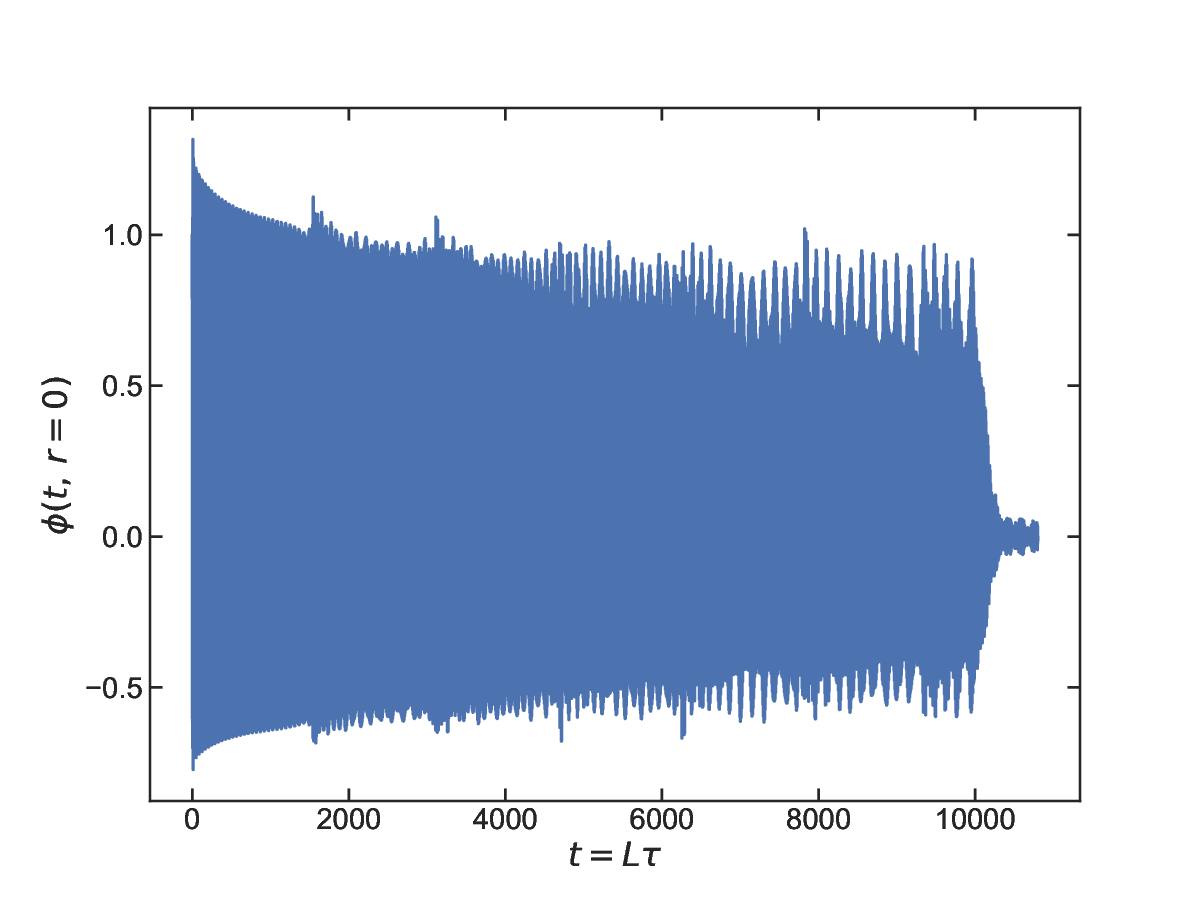}
    \caption{Time evolution of $\phi(t,\,r=0)$\,.}
    \label{fig1}
\end{figure}

\begin{figure}[htbp]
    \centering
    \includegraphics[width=12 cm]{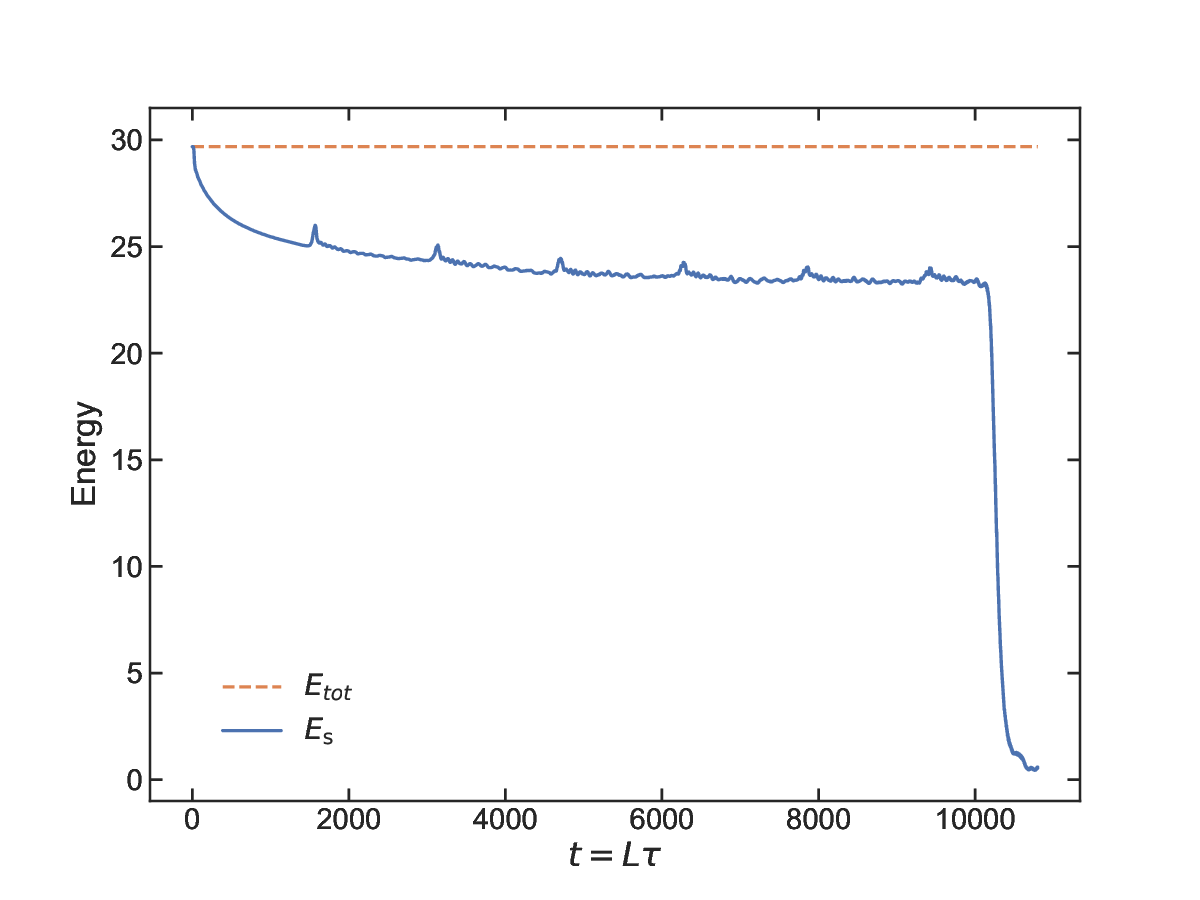}
    \caption{Time evolution of the total and shell energies.}
    \label{fig2}
\end{figure}

Figure \ref{fig1} shows the time evolution of the field $\phi$ at $r=0$\, with $\alpha=2.3$\,, $R_0=3.8$ and $L=500$\,. To see the consistency of the computation, the time evolution of the total and shell energies is presented in Fig.\,\ref{fig2}.  The orange line denotes the total energy $E_{\rm tot}$ in (\ref{E}) and the blue line indicates the shell energy $E_{\rm s}$ in (\ref{Es}) with $R_{\rm s} = 15$\,. The total energy $E_{\rm tot}$ is well conserved and ensures the consistency of the numerical computation. To be more precise, we have checked that numerical errors are $|(\,E_{\rm tot}(t) - E_{\rm tot}(t=0)\,)\,/\,E_{\rm tot}(t=0)|<10^{-4}$ at the end of computation. 

\medskip 

Seeing the behavior of $E_{\rm s}$\,, one can figure out the behavior of the solution. At first, a small amount of the energy escapes from the shell as radiation but most of $E_{\rm s}$ remains inside the shell. By combining this observation with Fig.\,\ref{fig1}\,, one can see that the excitation is well localized around $r=0$\,. Then, the field begins and continues to oscillate for a long time. This is a characteristic feature of the oscillon and is called the oscillon regime. Finally, the localized excitation decays around $t=10000$ and all remaining energy is abruptly emitted from the shell. This is a typical behavior of the oscillon as in the Minkowski case \cite{Copeland:1995fq}.

\medskip 

Note that there are some small bumps in the oscillon regime before the decay. These are radiations emitted at the beginning of time evolution and reflected by the AdS curvature effect. The traveling time is estimated as $t = \pi L$\,. Now it is $500\pi$ and nicely agrees with the periodicity in Fig.\,\ref{fig1}. This is a characteristic behavior intrinsic to the AdS space.

\subsection{Recurrence}

\begin{figure}[htbp]
    \centering
    \includegraphics[width=\columnwidth]{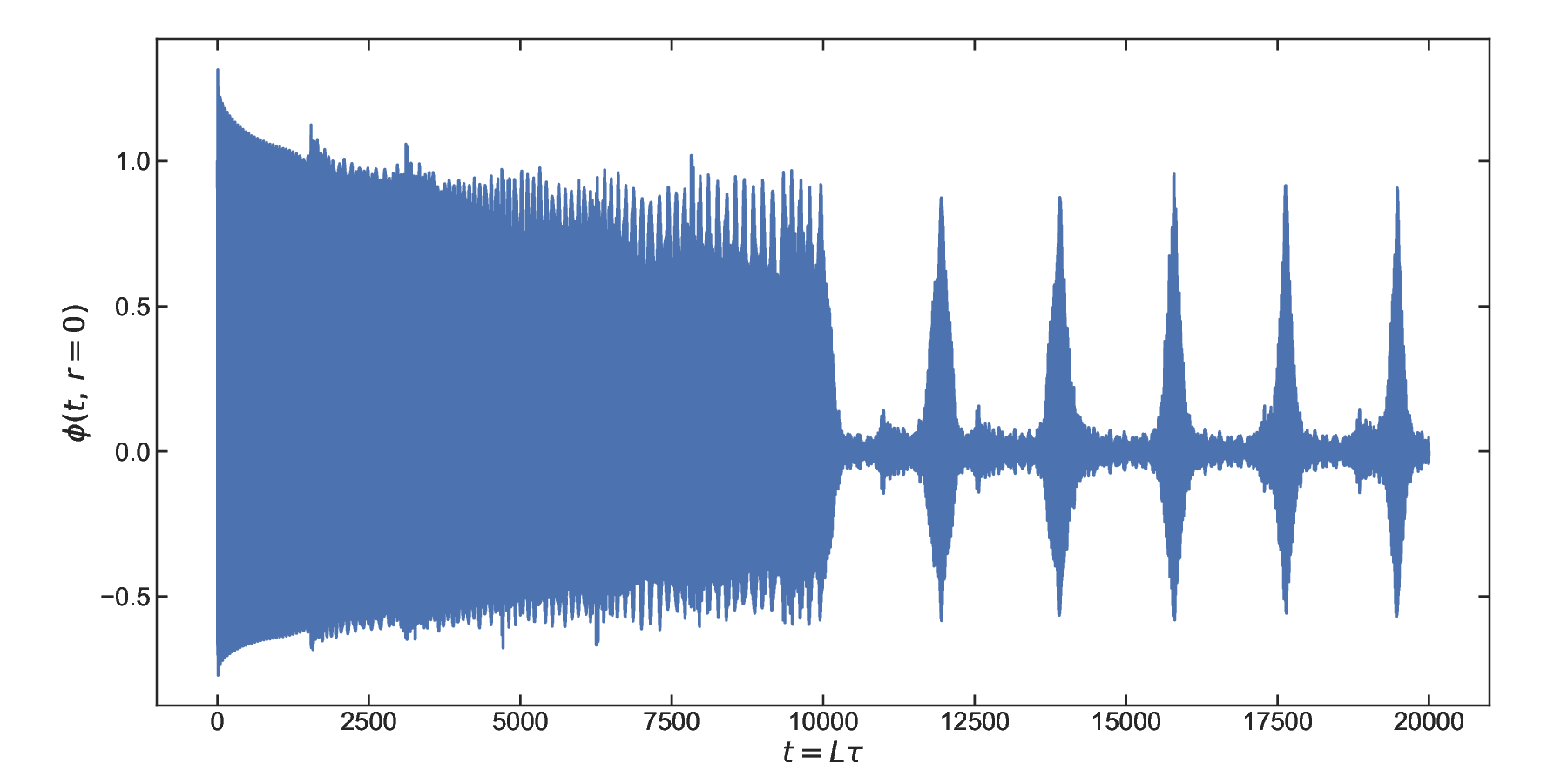}
    \caption{Time evolution of $\phi(t,\,r=0)$\,.}
    \label{fig3}
\end{figure}

\begin{figure}[htbp]
    \centering
    \includegraphics[width=\columnwidth]{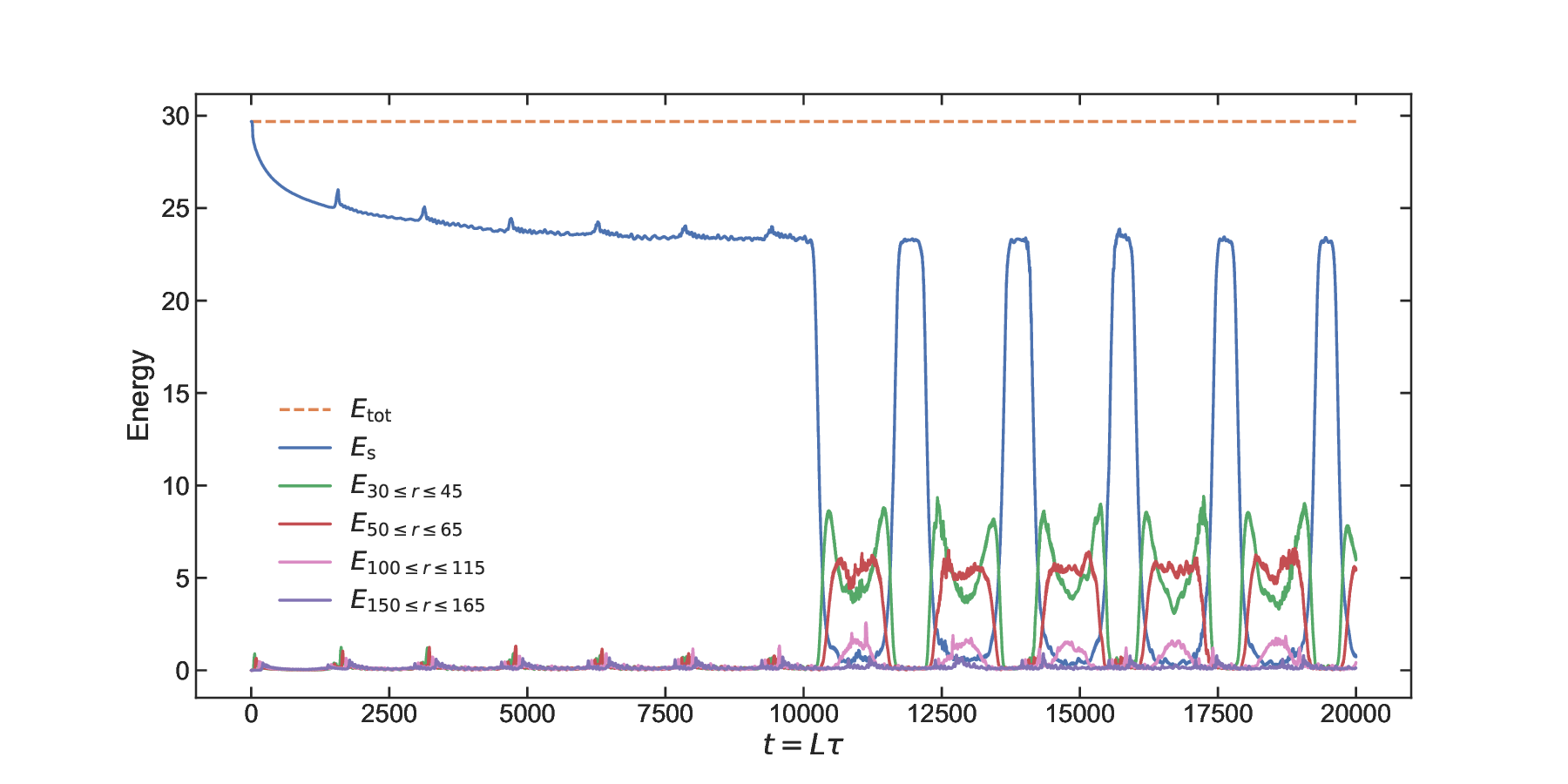}
    \caption{Time evolution of energies in different integral regions.}
    \label{fig4}
\end{figure}

Another characteristic of our setup can be seen by extending the computation time. 
The other parameters are the same as for 
Fig.\,\ref{fig1}. The result is plotted in Fig.\,\ref{fig3}. Now one can see new peaks after the end of the oscillon regime. To interpret them, it is useful to look at Fig.\,\ref{fig4}\,. By seeing the blue line, one can see that the energy released as radiation after the decay returns to $r=0$ periodically. The periodicity agrees with $500\pi$ again as small bumps in the oscillon regime. This can be understood as the phenomenon of recurrence.

\begin{figure}[htbp]
    \centering
    \includegraphics[width = 16cm]{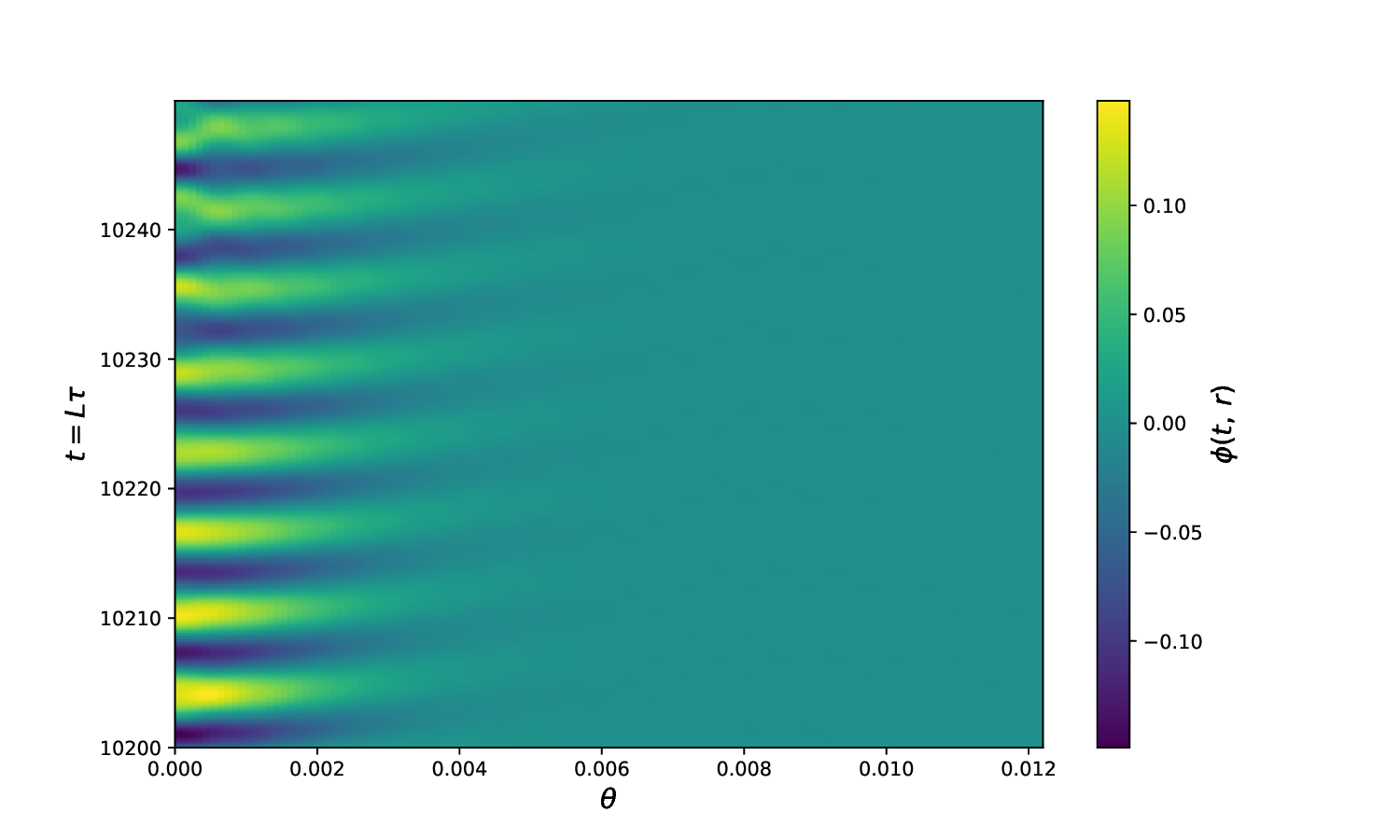}
    \caption{The oscillon behavior around the decay time.}
    \label{fig5}
\end{figure}

\medskip 

As a possibility, one might think that a localized lump begins to move without decay and then is reflected to $r=0$\,.  However, this can be excluded by seeing the green, red, pink and purple lines in Fig.\,\ref{fig4}\,. These denote the behavior of energies in different integral regions. Each energy $E_{a \leq r \leq b}(t)$ is defined as 
\begin{align}
    E_{a \leq r \leq b }(t) \equiv \int_{\theta_a}^{\theta_b}\!\!d\theta\,\frac{L}{\cos^2{\theta}}\,\mathcal{E}\left[\phi\right]\,, \quad 
    a \equiv L \tan\theta_{a}\,, \quad b \equiv L \tan\theta_{b}\,.
\end{align}
As a result, the behaviors indicate that there is no localized lump. This observation can also be seen from Fig.\,\ref{fig5}. This is a magnified plot around the oscillon decay. One can see that the oscillon emits radiation and tends to decay.  

\medskip 

{One may wonder if the value of $\lambda$ is sensitive to generate oscillons. In the above computation, we have set $\lambda=1$\,. In this case the system is strongly coupled and perturbative computations are not valid. While there is no problem at classical level, the oscillon may not be able to survive strong corrections at quantum level. Therefore, it is worth confirming that oscillons exist for small coupling constants for which perturbative computations are valid. This analysis has been done in Appendix B. }

\medskip

\begin{figure}[htbp]
   \centering
    \includegraphics[width=12cm]{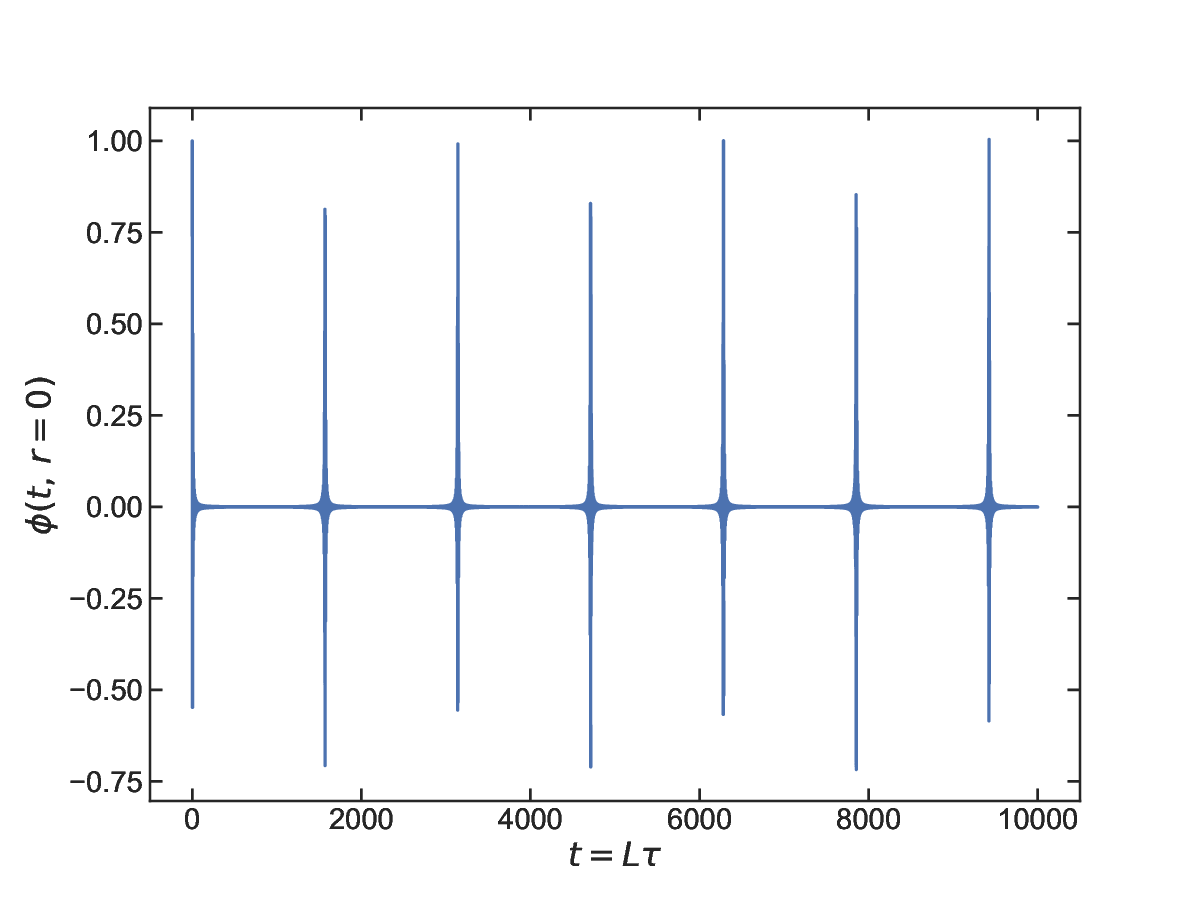}
    \caption{Time evolution of $\phi(t,\,r=0)\,$ in the non-oscillon case.}
    \label{fig6}
\end{figure}

\begin{figure}[htbp]
    \centering
    \includegraphics[width=12cm]{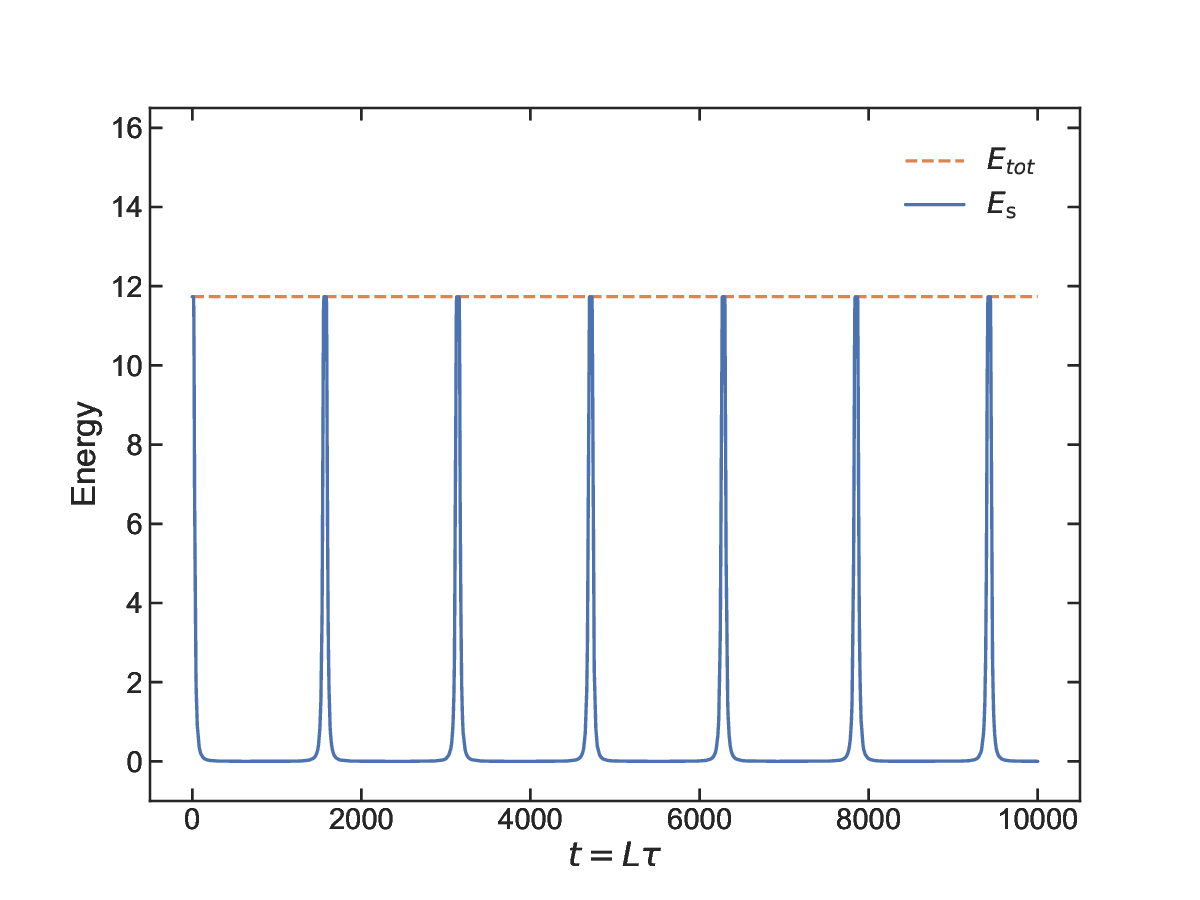}
    \caption{Time evolution of energies in the non-oscillon case.}
    \vspace*{0.5cm}
    \label{fig7}
\end{figure}

It is also interesting to see the recurrence phenomenon for non-oscillon configuration. 
In the Minkowski case, the bound for possible size of the oscillon core was evaluated in \cite{Copeland:1995fq}. This analysis can be generalized to the AdS case and the bound for the oscillon core size $R_0$ is given by\footnote{This is the condition for the case with $m^2>0$\,. In the AdS case, negative values of $m^2$ are possible if they satisfy the Breitenlohner-Freedman (BF) bound \cite{BF}. It is intriguing to study the oscillon in the regime with $m^2<0$\,. We will leave it as a future issue.} 
\begin{align}
    R_0^2\,>\,\frac{d}{\dfrac{\alpha^2}{3}\left(\dfrac{2\sqrt{2}}{3}\right)^d-1-\dfrac{d(d+2)}{4L^2}} 
    \label{cond}
\end{align}
and this condition provides a necessary condition for the existence of oscillon. {(See Appendix C for details of derivation and \eqref{bound1} for the final expression.)} By taking the $L\to \infty$ limit, the Minkowski result \cite{Copeland:1995fq} is reproduced. When $d=3,\,\alpha=2.3$ and $L=500$\,, the bound becomes $R_0>2.51$\,. In the previous computation, we took $R_0=3.8$ and the condition (\ref{cond}) was satisfied. By taking a value of $R_0$\,, one can generate non-oscillon solution. Figures \ref{fig6} and \ref{fig7} show the results for non-oscillons $R_0=2.4$. One can see that the localized lump decays immediately and all energy can be emitted as radiations. Then the radiations are reflected to $r=0$ periodically with period $500 \pi$\,. Thus, the recurrence can still be observed for non-oscillons and hence is not intrinsic to the oscillons. 

\medskip 

Finally, it should be mentioned that the phenomenon of the recurrence is rather surprising, because the system we are considering is not integrable even if the AdS space can be seen as a box. A typically expected behavior should be turbulent. At least so far, we have no idea to explain why the recurrence occurs even for non-oscillon configurations. It is the most significant issue to reveal the fundamental mechanism for the recurrence as well as the longevity of the oscillon. 

\subsection{$L$-dependence of the oscillon longevity}

So far, we have worked with a specific value of the AdS radius $L=500$\,. The reason we took it is to suppress the effect of the reflected waves. Since the period of the reflection is given by $\pi L$\,, it becomes shorter as the value of $L$ is smaller. In particular, if the value of $L$ is so small, the recurrence period also becomes so short that the reflected wave may be involved in the decaying wave packet because the decay process also takes some time. 

\medskip 

\begin{figure}[htbp]
\vspace*{0.3cm}
  \centering
  \begin{minipage}[b]{0.45\linewidth}
    \centering
    \includegraphics[width=\linewidth]{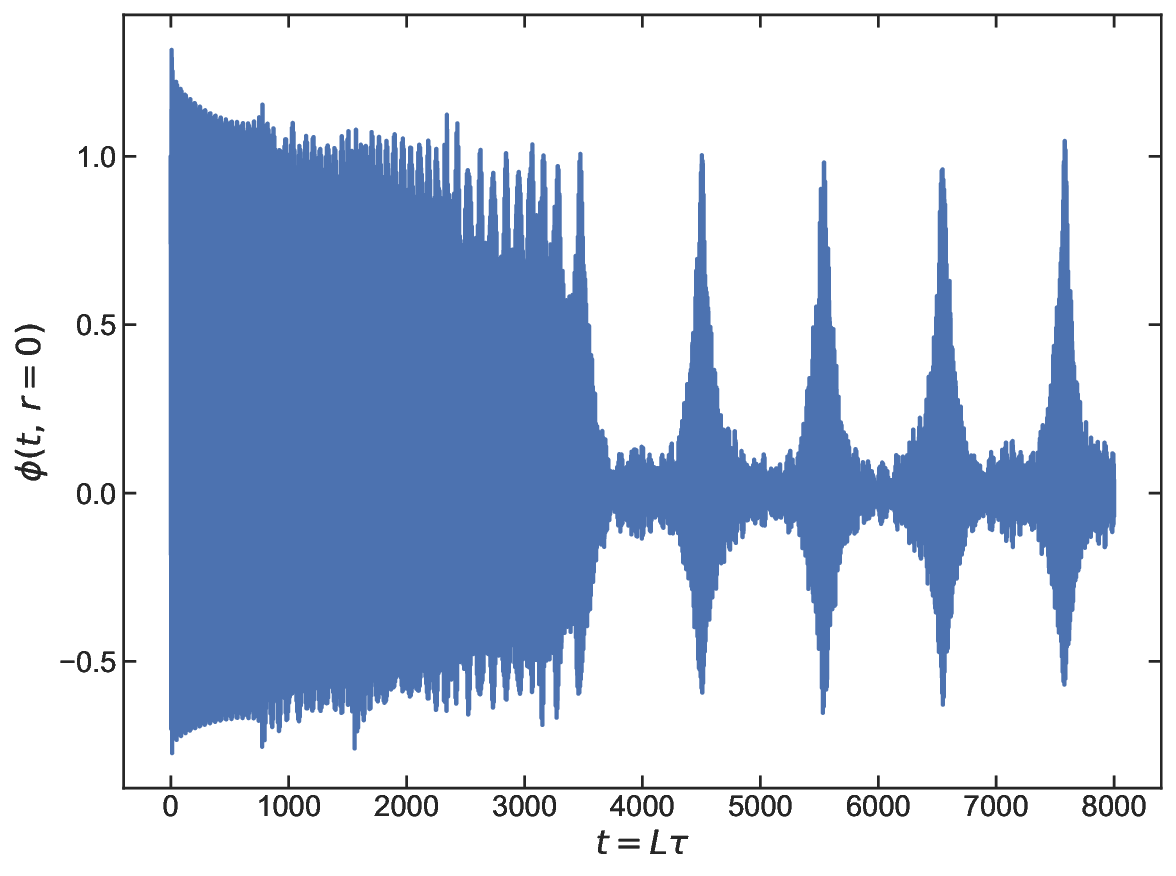}
    \label{fig:Fig8-1}
  \end{minipage}
  \hspace{0.05\linewidth}
  \begin{minipage}[b]{0.45\linewidth}
    \centering
    \includegraphics[width=\linewidth]{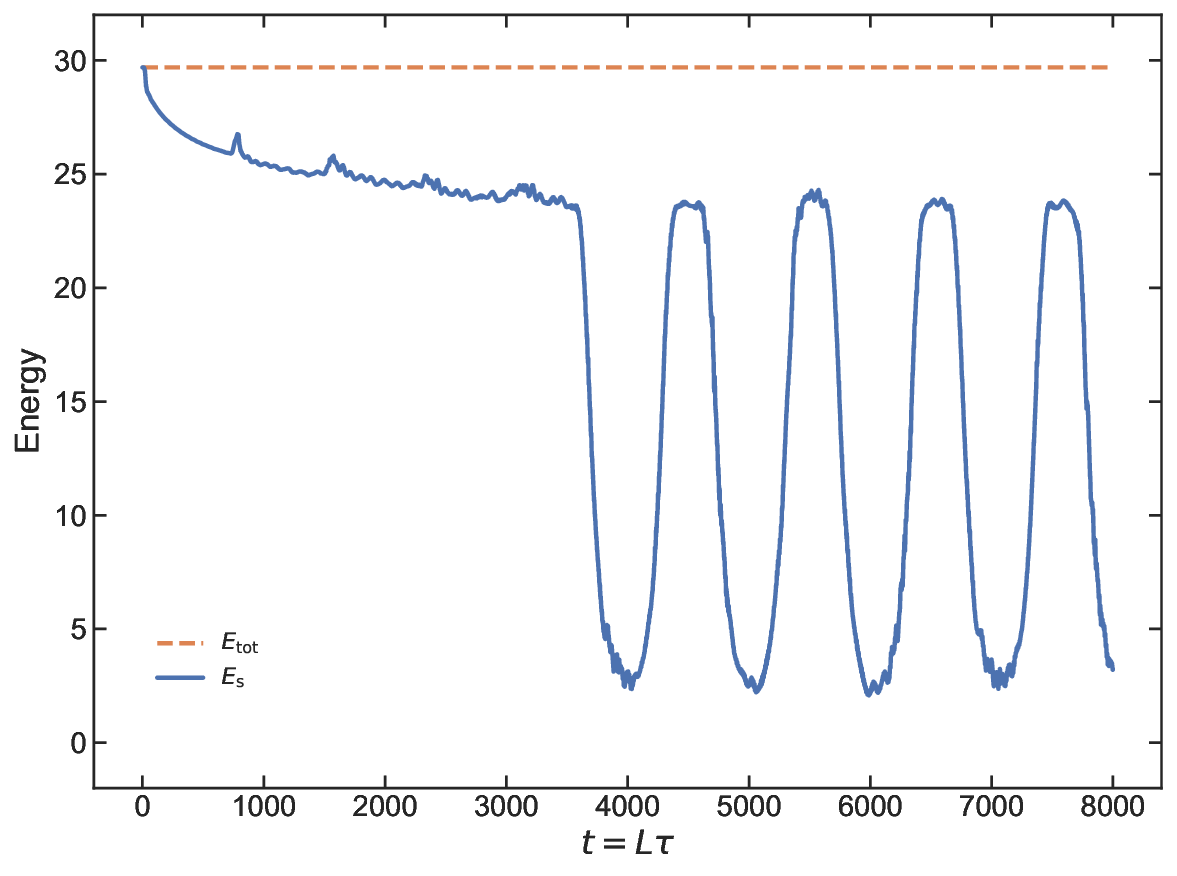}
    \label{fig:Fig8-2}
  \end{minipage}
  \vspace*{-0.5cm}
  \caption{Time evolution of $\phi(t,\,r=0)$ and the shell energy for $L=250$} 
  \label{fig8}
\end{figure}

Such a situation occurs, for example, with $L=250$ as presented in Fig.\,\ref{fig8}. The radiation emitted during the decay process returns back into the shell before completing the decay. Hence the shell energy does not become zero absolutely in comparison to the oscillon with $L=500$\,. 
The lifetime is around 3800 and the recurrent wave packets can still be seen sharply.

\medskip 

\begin{figure}[htbp]
\vspace*{0.3cm}
  \centering
  \begin{minipage}[b]{0.45\linewidth}
    \centering
    \includegraphics[width=\linewidth]{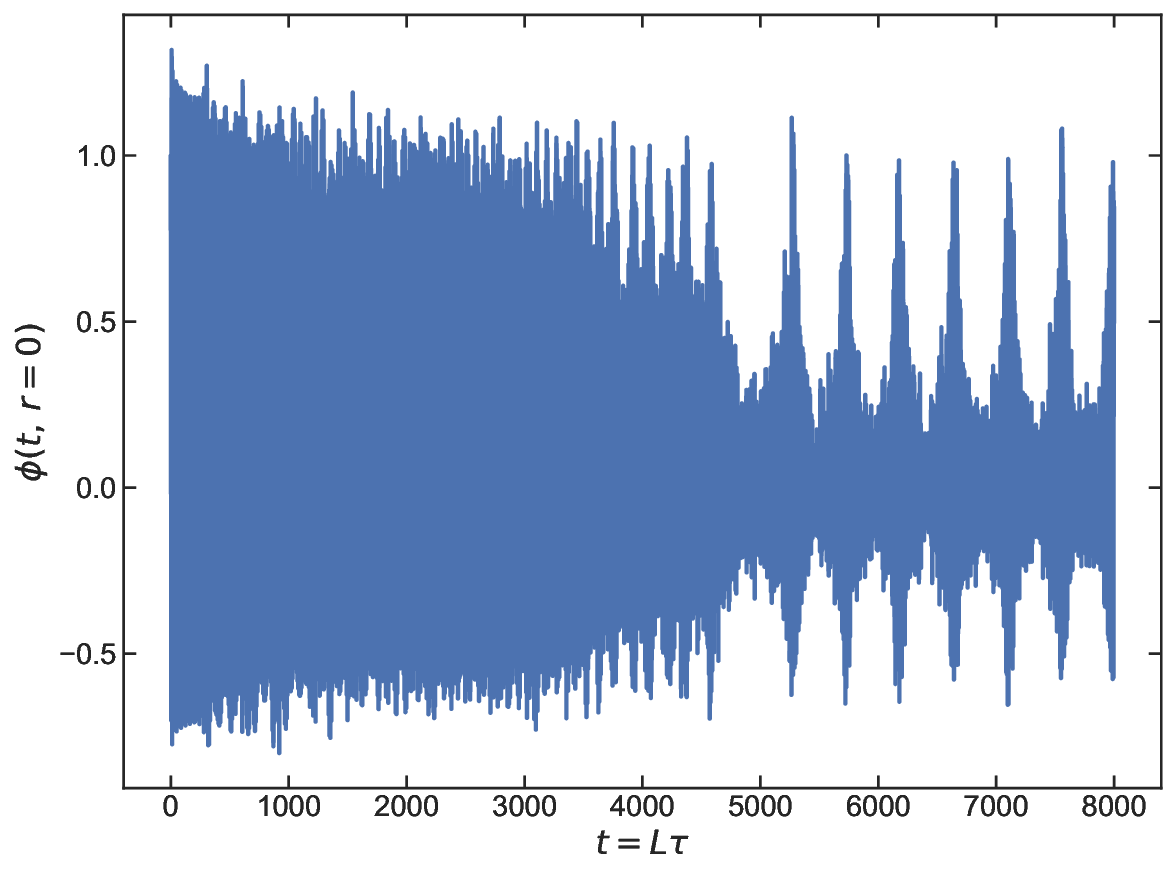}
    \label{fig:Fig9-1}
  \end{minipage}
  \hspace{0.05\linewidth}
  \begin{minipage}[b]{0.45\linewidth}
    \centering
    \includegraphics[width=\linewidth]{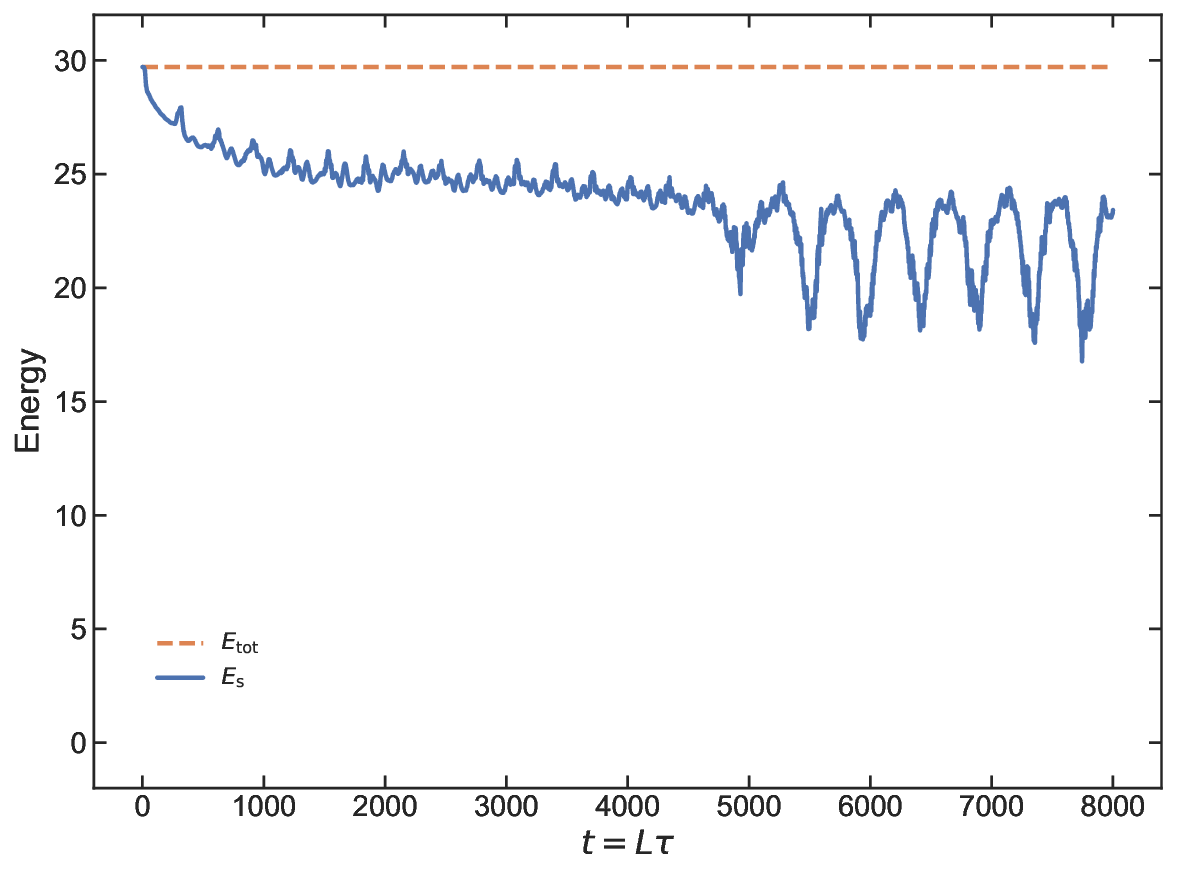}
    \label{fig:Fig9-2}
  \end{minipage}
    \vspace*{-0.5cm}
  \caption{Time evolution of $\phi(t,\,r=0)$ and the shell energy for $L=100$}
  \label{fig9}
\end{figure}

This tendency becomes even more pronounced when $L=100$ as displayed in Fig.\,\ref{fig9}. The reflected waves travel back to the shell much faster. Then the shell energy does not decrease drastically, but only shows small periodic gaps. The lifetime of the recurrent wave packet also becomes shorter.  

\begin{figure}[htbp]
\vspace*{0.3cm}
  \centering
  \begin{minipage}[b]{0.45\linewidth}
    \centering
    \includegraphics[width=\linewidth]{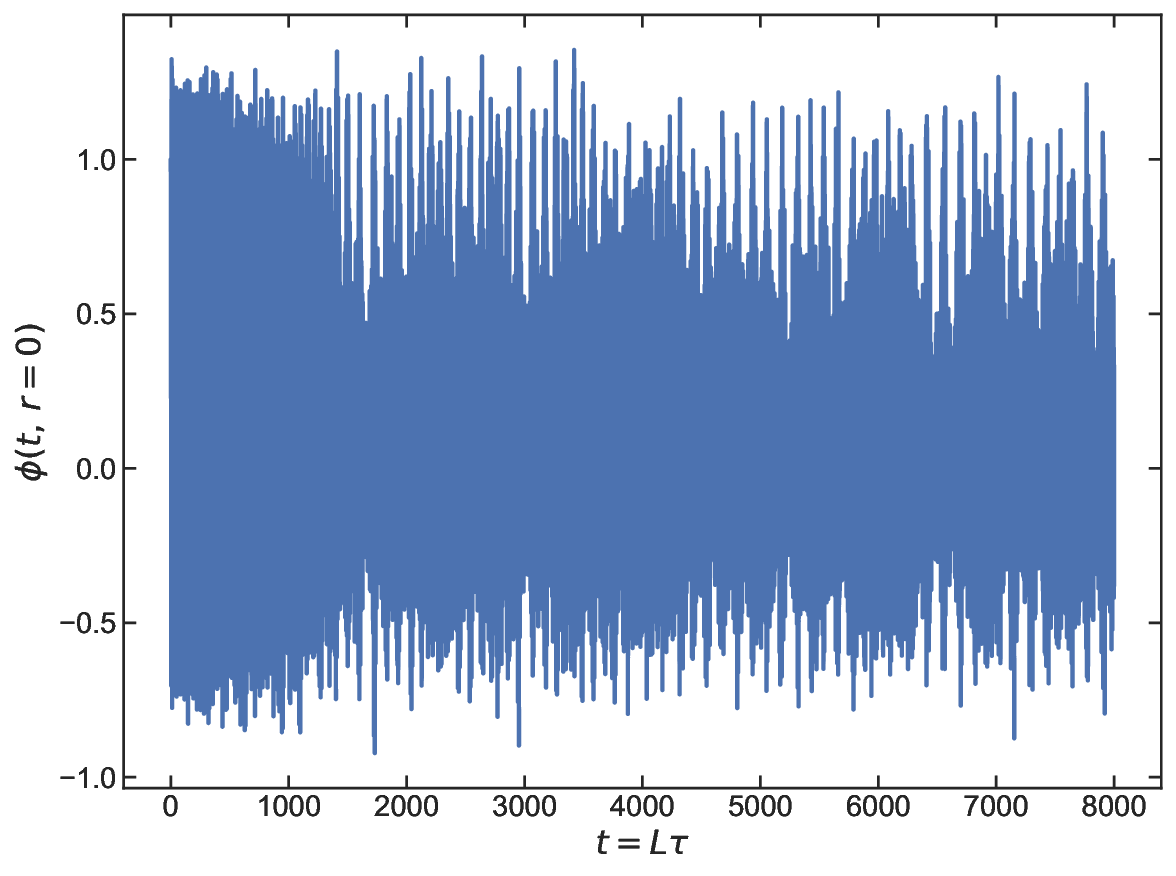}
    \label{fig:Fig10-1}
  \end{minipage}
  \hspace{0.05\linewidth}
  \begin{minipage}[b]{0.45\linewidth}
    \centering
    \includegraphics[width=\linewidth]{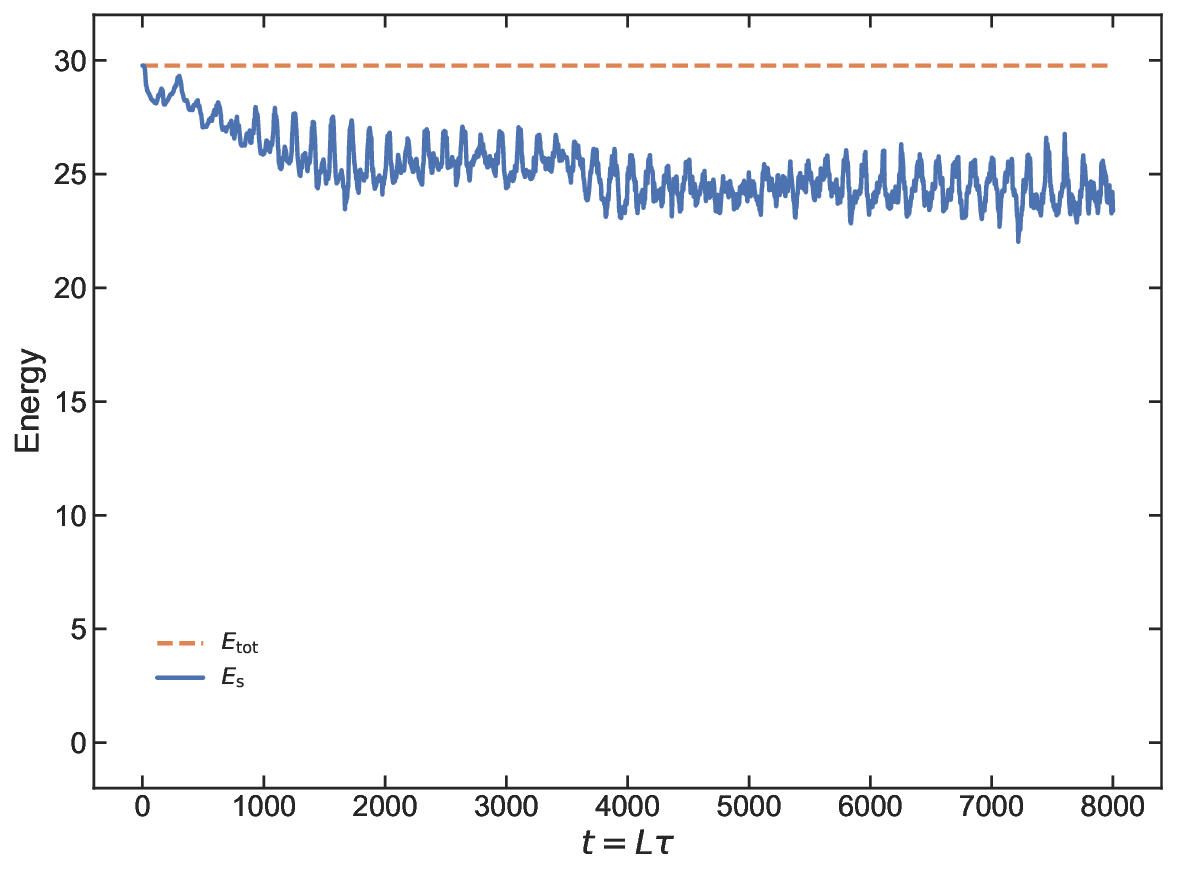}
    \label{fig:Fig10-2}
  \end{minipage}
    \vspace*{-0.5cm}
  \caption{Time evolution of $\phi(t,\,r=0)$ and the shell energy for $L=50$}
  \label{fig10}
\end{figure}

In the case with $L=50$\,, the decay process itself is severely disrupted by the reflected waves as shown in Fig.\,\ref{fig10}. The small periodic gaps in the shell energy can now be viewed as simple oscillations. As a consequence, oscillation at the origin is effectively stabilized. This can be seen as a kind of transition from a metastable oscillon to a stable oscillatory solution. This is an intriguing phenomenon intrinsic to the AdS case.

\medskip 

On the other hand, as derived in Appendix C, possible values of the AdS radius $L$ are bounded from below. Hence, it is interesting to investigate the oscillon behavior around the bound. For the parameters $d=3$\,, $\alpha=2.3,$ and $R_0 =3.8$
utilized to produce the above figures, the minimum of $L$ is evaluated as $L_{\rm min} = 3.73$\,. Hence $L=50$ satisfies the bound but is close to $L_{\rm min}$\,. 

\medskip 

As another note, the tail of the initial Gaussian configuration extends to around $r=10$\,. This size is estimated to be approximately $3R_0$\,, currently a value of the order of 10. The collapse of the wave packet at the origin is intensely suppressed by the reflected waves when $L$ is comparable to $3R_0$\,. As a consequence, it is quite significant to take the value of $L$ much larger than $3R_0$ so as to see the recurrence phenomenon clearly. On the other hand, when the value of $L$ is comparable to $3R_0$\,, the transition from a metastable oscillon to a stable oscillatory solution occurs.

\section{Conclusion and Discussion} 

We have studied oscillons in a real scalar field theory in a (3+1)-dimensional AdS space with the global coordinates. By supposing the same conditions as in the case of Minkowski spacetime, such as the Gaussian shape, we have constructed numerical oscillon solutions with longevity. In particular, since the AdS space can be seen as a box, the recurrence phenomenon has been observed under suitable conditions. {We have also found that an oscillon in AdS space shows a transition to a stable oscillatory solution when the AdS radius is small.}

\medskip

It may be interesting to consider an extension of the oscillon solution by including the gravitational backreaction for the AdS space by following the analysis \cite{BR}. The time-periodic solutions are constructed in \cite{MR} (for a nice review, see \cite{review}). The truly localized and time-periodic solutions in the AdS space without backreaction have already been discussed in \cite{FFG}. 

\medskip 

It is a fascinating issue to consider the dual gauge-theory interpretation of the oscillons (and extended ones with back reactions) via the AdS/CFT correspondence \cite{AdS-CFT}. There are some speculations and open problems. First, it may be interesting to consider a static string in an AdS space as discussed in \cite{DGT,GGK,SY1}. In this setup, real massive scalar field theories are realized in a two-dimensional AdS space and non-trivial potentials also appear by keeping some corrections in the semi-classical approximation. It is an intriguing challenge to look for an oscillon solution in this system. If it exists, it should be dual to a long-lived fluctuation of a Wilson loop \cite{Wilson1,Wilson2} on the gauge-theory side \cite{SY2}. 

\medskip

As a generalization in this direction, it is also possible to consider an AdS-brane in a higher-dimensional AdS space, which is a D-brane whose shape is AdS (called the AdS-brane). Similarly, real massive scalar field theories with a non-trivial potential can also be realized on the AdS-brane as well \cite{SY1}. There may also exist oscillon-like excitations on the brane and these should be dual to non-trivial fluctuations of the dual object such as (dual) giant Wilson loops \cite{giant0,giant1,giant2,giant3} and conformal defects \cite{defect}. 

\medskip 

As an application of the oscillons in the AdS space, it is nice to consider the AdS/QCD setup as discussed in \cite{QCD1,QCD2}. In this direction, we first need to construct an oscillon in Poincare AdS space. Then it is also necessary to somehow engineer the dilaton potential to support it.  
If such a oscillon could be realized, its gauge-theory dual would be related to the dynamics of a glueball-like excitation. This is a fascinating and ambitious research subject. 

\medskip 

The most important question is what is the underlying mechanism that ensures the oscillon longevity. Nobody knows the answer, at least for now. There may be a hidden symmetry behind the oscillon longevity, such as the Laplace-Runge-Lenz vector.  In the presence of mass terms, the free part is regarded as a collection of harmonic oscillators, and such a symmetry may approximately survive. In this direction, the work \cite{Evnin} may be relevant. 

\medskip 

Also, recently, there was an interesting approach to try to explain the oscillon longevity from the viewpoint of sphaleron \cite{sphaleron} in a real scalar field theory with a cubic potential in two-dimensional Minkowski spacetime \cite{Manton}. It is interesting to generalize the work \cite{Manton} to a two-dimensional AdS space. 

\medskip 

The mechanism to ensure oscillon longevity might be enlightened by the AdS/CFT correspondence. We hope that our work will be a first step in this direction.

\subsection*{Acknowledgments}

The authors thank Henry Liao for useful discussions and comments. 
The work of T.~I.\ was supported by JSPS Grant-in-Aid for Scientific Research (C) No.\,19K03871. The works of T.~M.\ and K.~Y.\ were supported by MEXT KAKENHI Grant-in-Aid for Transformative Research Areas A ``Machine Learning Physics'' No.\,22H05115, and JSPS Grant-in-Aid for Scientific Research (B) No.\,22H01217 and Scientific Research (C) No.\, 25K07313.

\appendix

\section*{Appendix}

\section{Rescaling of the Lagrangian}

To perform numerical computations, let us rewrite the Lagrangian in \eqref{action}\,, 
\begin{align}
    \mathcal{L}
    =
    V_{d-1} \,r^{d-1}
    \left[
    -\frac{1}{2}g^{tt}
    \partial_t\phi\partial_t\phi
    -\frac{1}{2}g^{rr}
    \partial_r\phi\partial_r\phi
    -\frac{m^2}{2}\phi^2
    +\frac{\alpha_0}{3}\phi^3
    -\frac{\lambda}{4}\phi^4
    \right]\,
    \label{original}
\end{align}
into the dimensionless form. 

\medskip 

First of all, the coordinates $t$ and $r$ are made dimensionless by using the mass $m$ as
\begin{align}
    \hat{t} \equiv m\,t\,, \qquad  \hat{r} \equiv m\,r\,,
    \label{rescaling-coordinates}
\end{align}
where $\hat{t}$ and $\hat{r}$ are dimensionless time and radial coordinates. Note here that this rescaling is intrinsic to the massive case. 

\medskip 

Then, the curvature radius $\ell$ and the coupling constants $\alpha_0$ and $\lambda$ are also rescaled as 
\begin{align}
        L \equiv m\,\ell\,, \qquad  
    \hat{\alpha}_0 \equiv \frac{1}{m^{(5-d)/2}}\,\alpha_0\,, \qquad
    \hat{\lambda} \equiv \frac{1}{m^{3-d}}\,\lambda\,,
    \label{rescaling-constants}
\end{align}
and the dimensionless field is defined as 
\begin{align}
    \hat{\phi} \equiv
    \frac{\sqrt{\hat{\lambda}}}
    {m^{(d-1)/2}}\,\phi\,.
    \label{rescaling-field}
\end{align}
Finally, by adjusting the overall constant, the dimensionless Lagrangian is given by 
\begin{align}
    \mathcal{L}_\mathrm{rescaled}
    =
    \frac{1}{\hat{\lambda}}\,V_{d-1} \,\hat{r}^{d-1}
    \left[
    -\frac{1}{2}\hat{g}^{\hat{t}\hat{t}}
    \partial_{\hat t}\hat\phi\partial_{\hat t}\hat\phi
    -\frac{1}{2}\hat{g}^{\hat{r}\hat{r}}
    \partial_{\hat r}\hat\phi\partial_{\hat r}\hat\phi
    -\frac{1}{2}\hat{\phi}^2
    +\frac{\alpha}{3}\,\hat{\phi}^3
    -\frac{1}{4}\hat{\phi}^4
    \right]\,, 
    \label{dimless}
\end{align}
where $\alpha \equiv \hat{\alpha}_0\,/\,\sqrt{\hat{\lambda}}$\,. 

\medskip 

In the main text, all the hats have been removed for brevity.

\section{Towards including quantum corrections}

In the main text, we have taken the rescaling described in Appendix A, where the dimensionless quartic coupling is set to 1. This is convenient for our classical analysis. However, if we consider at the quantum level, large quantum corrections would destroy the classical result because the perturbative approximation is not valid anymore. 
Therefore, it is significant to present oscillons with the parameters for which the perturbative approximation is valid. 

\medskip 

Let us rescale the coordinates and the curvature radius in the same manner as in  (\ref{rescaling-coordinates}) and (\ref{rescaling-constants}), and rescale the coupling constants and the field as 
\begin{align}
    \tilde{\alpha}_0
    \equiv
    \frac{2}{m^{(5-d)/2}}\,\alpha_0\,, \qquad
    \tilde{\lambda}
    \equiv
    \frac{6}{m^{3-d}}\,\lambda\,, \qquad
    \tilde\phi
    \equiv
    \frac{1}{m^{(d-1)/2}}\phi\,. 
\end{align}
Then the dimensionless potential takes the form
\begin{align}
    V(\phi)
    =
    \frac{1}{2}\phi^2 - \frac{\alpha_0}{3!}\phi^3 + \frac{\lambda}{4!}\phi^4\,, 
    \label{C2}
\end{align}
where the hats and tildes have been removed. Here, in comparison to the potential (\ref{potential})\,, the coefficients of the cubic and quartic potential have been modified to adopt the usual field theory convention. This is relevant to the values of the coupling constants.

\begin{figure}[htbp]
\vspace*{0.3cm}
  \centering
  \begin{minipage}[b]{0.45\linewidth}
    \centering
    \includegraphics[width=\linewidth]{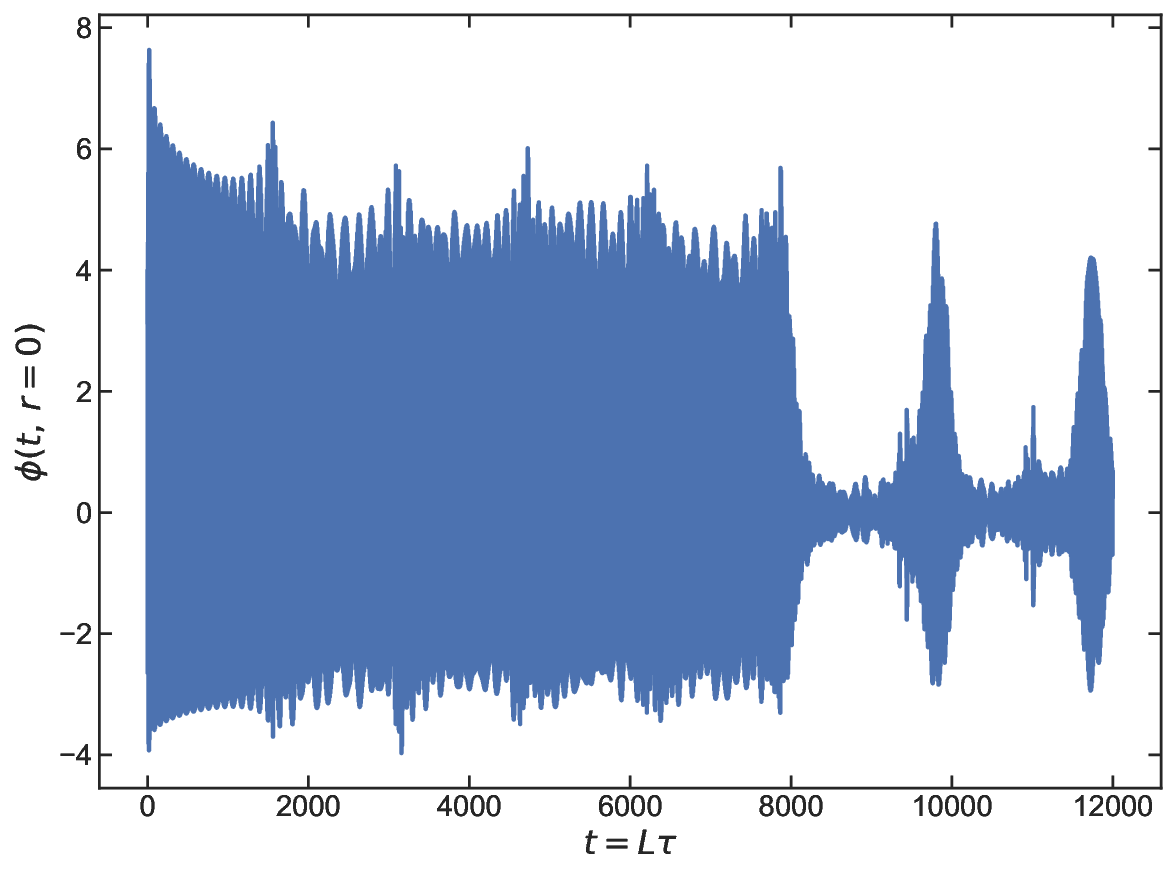}
    \label{fig:Fig11-1}
  \end{minipage}
  \hspace{0.05\linewidth}
  \begin{minipage}[b]{0.45\linewidth}
    \centering
    \includegraphics[width=\linewidth]{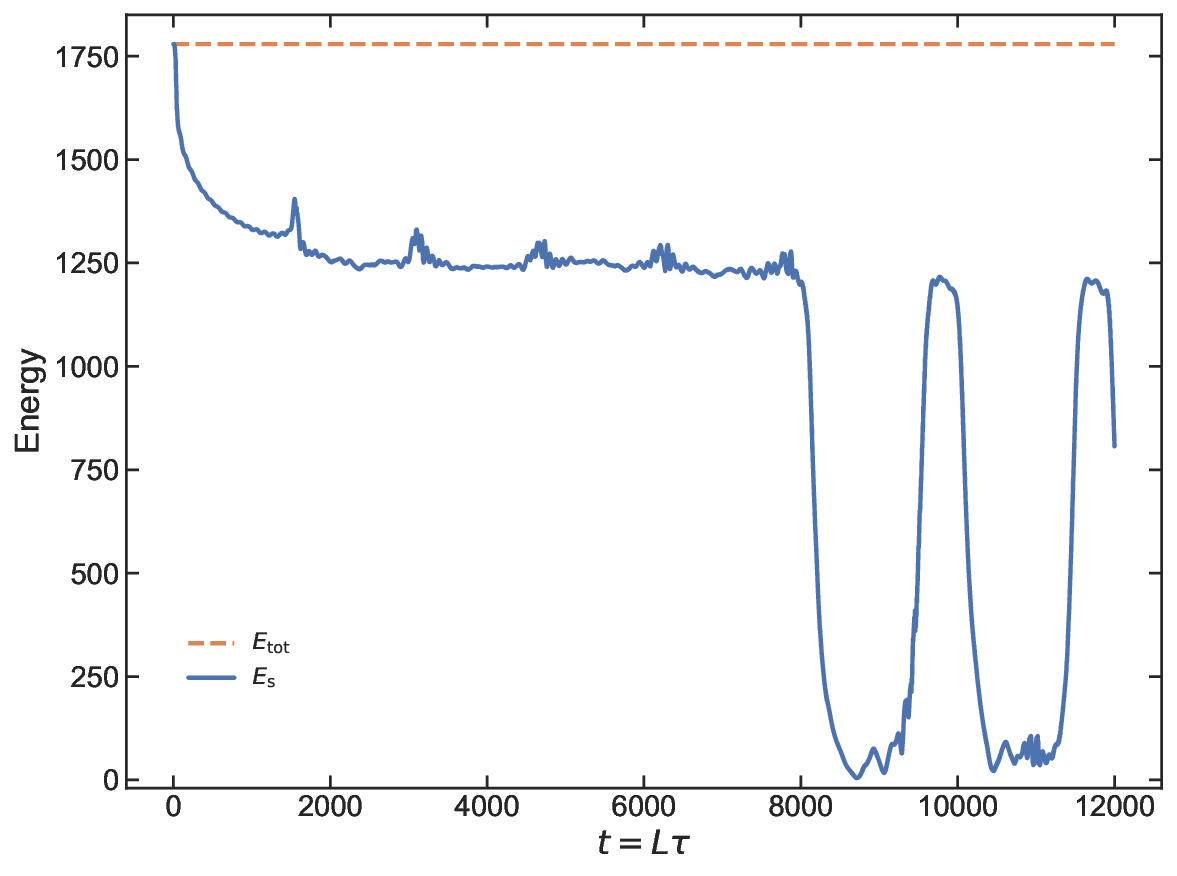}
    \label{fig:Fig11-2}
  \end{minipage}
  \vspace*{-0.5cm}
  \caption{Oscillon with small $\alpha$ and $\lambda$\,. Time evolution of $\phi(t,\,r=0)$ and the shell energy are displayed on the left and right sides, respectively.} 
  \label{fig11}
  \vspace*{0.5cm}
\end{figure}

\medskip 

We present oscillons in the perturbative regime by using the potential (\ref{C2}) and the Gaussian initial configuration
\begin{align}
     &\phi(t=0,\,r) = A\,{\rm e}^{-r^2/R_0^2}\,,
\end{align}
where an initial amplitude $A$ is a real constant. Numerical results are displayed in Fig.\,\ref{fig11}, where the time evolution of the field at $r=0$ (the left-hand side) and the shell energy (on the right-hand side) with parameters $\alpha_0=0.7$, $\lambda=0.18$, $A=4.0$, $R_0=5.51$ and $L=500$\,. The lifetime is around 8000 and the recurrence phenomenon can still be observed, though the value of $L$ is taken to be large so as to see the recurrence clearly. 

\medskip

Thus, we have succeeded in constructing an oscillon with parameters for which perturbative computations are valid.

\section{A bound on the oscillon core size}

In the case of Minkowski spacetime, a possible bound for the size of the oscillon core has been discussed in \cite{Copeland:1995fq}. Following the analysis in \cite{Copeland:1995fq}, we shall examine a bound on the size of the oscillon core for $m^2 > 0$ in AdS space. 

\medskip
 
Following \cite{Copeland:1995fq}, let us first suppose the following oscillon profile:
\begin{align}
    \phi(t,\,r) = q(t)\,e^{-r^2/R_0^2}\,.
    \label{oscillon_model}
\end{align}
Here, $q(t)$ is an oscillating function, 
and $R_0$ denotes the initial core size. This profile is motivated by numerical observations \cite{Copeland:1995fq} indicating that the oscillon maintains the Gaussian shape before the decay. 

\medskip 

By applying the profile (\ref{oscillon_model}) to our setup\footnote{Here, we have restored $\lambda$\,. The setting in the main text is recovered by setting $\lambda=1$\,.}, we can see a necessary condition for the existence of an oscillon in the AdS space. Substituting \eqref{oscillon_model} into \eqref{Lagrngian_rescaled} leads to the following effective Lagrangian: 
\begin{align}
    \frac{\lambda}{A}\,L_\mathrm{eff} = \frac{1}{2}\dot{q}^2 - \frac{C}{2A}\left(1+\frac{B}{C}\right)q^2 + \frac{\alpha}{3}\frac{D}{A}q^3 - \frac{1}{4}\frac{F}{A}q^4\,,
\end{align}
where the quantities $A$, $B$, $C$, $D$ and $F$ are defined as 
\begin{align}
    A &\equiv \frac{V_{d-1}}{\dot{q}^2}\int\! dr\,r^{d-1}
    \,f^{-1}\,
    \!\dot{\phi}^2 = \pi^{d/2}\,e^{2L^2/R^2}\,L^d\,\Gamma\!\left(1-\frac{d}{2},\,\frac{2L^2}{R_0^2}\right)\,,\label{quantity-A}\\
    B &\equiv \frac{V_{d-1}}{q^2}\int\! dr\,r^{d-1}
    \,f\,
    {\phi^\prime}^2 = \frac{d\pi^{d/2}(4L^2+(2+d)R_0^2)}{2^{2+d/2}}\frac{R_0^{d-2}}{L^2}\,,\label{quantity-B}\\
    C &\equiv \frac{V_{d-1}}{q^2}\int\! dr\,r^{d-1}\phi^2 = \frac{\pi^{d/2} R_0^d}{2^{d/2}}\,,\label{quantity-C}\\
    D &\equiv \frac{V_{d-1}}{q^3}\int\! dr\,r^{d-1}\phi^3 = \frac{\pi^{d/2}R_0^d}{3^{d/2}}\,,\label{quantity-D}\\
    F &\equiv \frac{V_{d-1}}{q^4}\int\! dr\,r^{d-1}\phi^4 = \frac{\pi^{d/2}R_0^d}{2^d}\,,\label{quantity-F}
\end{align}
where $\Gamma(s,x)$ denotes the upper incomplete gamma function. 
Then $q(t)$ obeys a non-linear ordinary differential equation,
\begin{align}
    \ddot{q} = -\frac{C}{A}\left(1 + \frac{B}{C}\right)q + \alpha\frac{D}{A}q^2 -  \frac{F}{A}q^3\,. \label{diff}
\end{align}
Suppose that $\bar{q}(t)$ is a solution to (\ref{diff})\,. Then, by expressing $q(t)$ around the solution $\bar{q}(t)$ with a fluctuation $\delta q(t)$ 
like $q= \bar{q}+\delta q$\,, we can obtain a linearized equation for $\delta q$\,,
\begin{align}
    \ddot{\delta q} = - \omega^2(\bar{q})\,\delta q\,,
    \label{fluctuation eq}
\end{align}
where the square of the frequency is given by  
\begin{align}
    \omega^2(\bar{q})\equiv 3\frac{F}{A}\bar{q}^2 - 2\alpha\frac{D}{A}\bar{q} + \frac{C}{A}\left(1+\frac{B}{C}\right).
\end{align}
When $\omega^2$ is positive, the equation \eqref{fluctuation eq} implies that the solution $\bar{q}$ is stable. If we consider the radiation of the energy, which is not included in the model (\ref{oscillon_model}), the stable solution decays into the vacuum $q=0$\,. 

\medskip

On the other hand, when $\omega^2$ is negative, the solution to \eqref{fluctuation eq} becomes unstable and drives the amplitude $q(t)$ away from $\bar{q}$\,. This instability is essential for the oscillon to appear. Hence, $\omega^2$ should be negative for a certain region of $\bar{q}$ and the discriminant must be positive as follows: 
\begin{align}
     \alpha^2\frac{D^2}{A^2} - 3\frac{CF}{A^2}\left(1 + \frac{B}{C}\right) > 0\,.
\end{align}
Substituting \eqref{quantity-A}, \eqref{quantity-B}, \eqref{quantity-C}, \eqref{quantity-D}, and \eqref{quantity-F} into this inequality, we obtain
\begin{align}
    \left[\dfrac{\alpha^2}{3}\left(\dfrac{2\sqrt{2}}{3}\right)^d-1-\dfrac{d(d+2)}{4L^2}\right]R_0^2\,>\,d\,.
    \label{instability-inequality}
\end{align}
This inequality provides a necessary condition for the existence of oscillons. Because the parenthesis in (\ref{instability-inequality}) must be positive, we find the following condition
\begin{equation} 
\dfrac{\alpha^2}{3}\left(\dfrac{2\sqrt{2}}{3}\right)^d-1 > \dfrac{d(d+2)}{4L^2}\,, 
\label{cond1}
\end{equation}
which is necessary to generate oscillons. If the parameters do not satisfy the above inequality, oscillons cannot be generated.  
Then,  under the condition (\ref{cond1})\,, the inequality \eqref{instability-inequality} can be seen as a lower bound on the oscillon core size $R_0$ in the AdS case:
\begin{align}
    R_0^2\,>\,\frac{d}{\dfrac{\alpha^2}{3}\left(\dfrac{2\sqrt{2}}{3}\right)^d-1-\dfrac{d(d+2)}{4L^2}}\,.
    \label{bound1}
\end{align}
In the limit $L\to \infty$\,, the bound in the Minkowski case  \cite{Copeland:1995fq} is reproduced. 

\medskip 

Intriguingly, the inequality (\ref{instability-inequality}) can be rewritten as 
\begin{align}
    \left[\frac{\alpha^2}{3}\left(\frac{2\sqrt{2}}{3}\right)^d - 1 - \frac{d}{R_0^2}\right]L^2 \,>\, \frac{d(d+2)}{4}\,,
    \label{instability-inequality2}
\end{align}
and the positivity inside the parenthesis gives the following inequality necessary for generating oscillons:
\begin{align}
    \frac{\alpha^2}{3}\left(\frac{2\sqrt{2}}{3}\right)^d - 1 > \frac{d}{R_0^2}\,.
\end{align}
When this inequality satisfied, the inequality (\ref{instability-inequality2}) means a lower bound on the AdS radius,
\begin{align}
    L^2\,>\,\frac{d(d+2)/4}{\dfrac{\alpha^2}{3}\left(\dfrac{2\sqrt{2}}{3}\right)^d-1-\dfrac{d}{R_0^2}}\,. \label{B4}
\end{align}

\medskip 

So far, we have discussed the bounds by using the Lagrangian \eqref{Lagrngian_rescaled}. It is also worth seeing the inequality for the original form of the Lagrangian (\ref{original}) by keeping the parameters explicitly. The derivation is straightforward, and hence we will not present the derivation here. The resulting bounds for $R_0^2$ and $\ell^2$ are given by, respectively,\footnote{Here we use the original curvature radius $\ell$ appearing in \eqref{adsds2} because it has not yet been rescaled in the Lagrangian \eqref{original}.}
\begin{align}
    R_0^2\,&>\,\frac{d}{\dfrac{\alpha^2}{3\lambda}\left(\dfrac{2\sqrt{2}}{3}\right)^d - m^2 -\dfrac{d(d+2)}{4\ell^2}}\,,\\
    \ell^2\,&>\,\frac{d(d+2)/4}{\dfrac{\alpha^2}{3\lambda}\left(\dfrac{2\sqrt{2}}{3}\right)^d - m^2 - \dfrac{d}{R_0^2}}\,. 
\end{align}

Note here that for the setup considered in Appendix B, a similar inequality can be derived, and the parameters used there satisfy the inequality.

\end{document}